# Induced magnetism by single carbon vacancies in a three-dimensional graphitic network: a supercell study

R. Faccio[1], H. Pardo[1], P. A. Denis[1], R. Yoshikawa Oeiras[2], F. M. Araújo-Moreira[2], M. Veríssimo-Alves[3] and A. W. Mombrú[1*]

[1] *Crystallography, Solid State and Materials Laboratory (Cryssmat-Lab), DETEMA, Facultad de Química, Universidad de la República, P.O. Box 1157, Montevideo, URUGUAY.*

[2] *Departamento de Física, Universidade Federal de São Carlos, CP 676, São Carlos, SP, Brazil, CEP 13565-905*

[3] *The Abdus Salam International Centre for Theoretical Physics, Strada Costiera 11, Main Building, Trieste I-34014, Italy*

## ABSTRACT

We present an *ab initio* DFT study of the magnetic moments that arise in graphite by creating single carbon vacancies in a 3-D graphite network, using a full potential, all electron, spin polarized electronic structure calculations. In previous reports the appearance of magnetic moments was explained in a 2-D graphene sheet just through the existence of the vacancies itself [1-5]. The dependence of the arising magnetic moment on the nature and geometry of the vacancies for different supercells is reported. We found that the highest value of magnetic moment is obtained for a 3x3x1 supercell and that the highly diluted 5x5x1 supercell shows no magnetic ordering. The results obtained in this manuscript are indicative of the importance of interlayer interactions present in a 3-D stacking. We conclude that this should not be underestimated when vacancies-based studies on magnetism in graphitic systems are carried out.

# I. INTRODUCTION

Carbon based materials have attracted the attention of the scientific community due to their many potential technological applications. For this reason, novel properties induced on graphite by chemical or physical modifications have become a subject of great interest [1-5]. Many theoretical articles focus not only on the defects of nanostructured materials, but in the contour edges or in the presence of other atoms [2-5]. In most of them the system was modeled as an interaction among atoms within a graphene (single graphite layer). In this report we are studying the effect of the presence of single atoms vacancies in a 3-dimensional graphitic network – bulk graphite-, with different supercells, with emphasis on the metallic nature and the magnetic response of this modified material.

# II. METHODOLOGY

The *ab initio* calculations were performed within the Density Functional Theory (DFT) framework, using a full potential all electrons, spin polarized electronic structure calculations utilizing the APW+lo method within the WIEN2k code [6]. For the exchange-correlation potential we used Generalized-Gradient-Approximation (GGA) in the Perdew-Burke-Ernzerhoff (PBE) scheme [7]. Scalar relativistic effects were included but spin-orbit coupling was neglected.

The convergence of the basis set is controlled by a cutoff parameter expressed as the product between the smallest muffin tin radius in the unit cell ($R_{MT}$) and the magnitude of the maximum reciprocal lattice vector ($K_{MAX}$). The muffin-tin radii for carbon atoms were selected as $R_{MT}$ = 1.25 a.u., achieving convergence for a cutoff value of $R_{MT}*K_{MAX}$= 5.5. The *1s* state was selected as a core state, while the *2s* and *2p* states were treated as valence states. In all the cases the valence charge density were expanded up to a $G_{MAX}$ value of 20.0 a.u.$^{-1}$, equivalent to kinetic energy $E_{cut}$= 400 Ry, in order to increase the accuracy in the determination of the topology of the electronic density. Finally, the k-points sampling in the first Brillouin Zone was determine specifically for each supercell (Table 1), studying convergence for different number of density points. In the case of pristine graphite we obtain a convergence in total energy with 16000 points in the entire first Brillouin Zone.

Layered materials present strong corrugation of the crystal potential in the direction perpendicular to the layers. Since this is the case of bulk graphite, the Local-Density-Approximation (LDA) exchange-correlation potential does not accurately reproduce the crystal potential of this system. For example, it is known that LDA tends to overestimate interlayer interactions in graphite, leading to slightly short distances along its *c* direction; or in some cases leading by chance close to the experimental one. For this reason, some reports support that the GGA xc-potential is more adequate in systems with important in-homogeneities on their charge densities along the direction perpendicular to the *c*-axis [8-10], and for the reconstruction in the a-b plane. They postulate the use of this potential instead of a LDA approach, since they find it more appropriate for carbon structures. Although the *Van der Waals* forces are totally neglected [11] for both xc-potentials, GGA leads to an increase of the interlayer distances, which causes that some authors prefers to use the LDA potential [5-10,12]. We present here the results of the supercell series using the GGA potential, but comparing the results extracted from calculations using LDA and GGA for the 3x3x1 supercell, in order to support the reliability of the conclusions presented in this report.

Graphite is a nonmagnetic semimetal that displays a very weak dispersion along *c*-axis. We started from the hexagonal graphite cell in the P6$_3$/mmc space group, A-B stacking, $a_0=b_0=4.656$ a.u. and $c_0=12.682$ a.u., and constructed a series of supercells. Then we removed single atoms on each graphene layer, in order to obtain vacancies aligned along *c*-axis, the $(0,0,¼)$ and $(0,0,¾)$, in different concentrations by means of different volumes of the supercells [13]. After this, we performed a spin polarized calculation with geometrical relaxation of the atomic positions. This last step was carried out minimizing the forces in the atoms and the total energy of the system. In all the cases the force tolerance was selected as 1.0 mRy/a.u.. This process was repeated for different *c* values until finding the one that minimizes the total energy; this is called the equilibrium result $c_{eq}$. In order to check that the ground state of the systems is magnetic we forced the system to converge to a non-magnetic solution, as in pristine graphite, and we found in all the cases a solution with higher energy than in the magnetic one (except for the 5x5x1 supercell, as will be shown below).

The localized states along the $c$ axis (see Figs. 2 and 3), could be related to the possible spontaneous magnetism of the material in agreement with previous reports [4-5]. The relaxation of the atoms within each supercell is an important effect that was taken into account in this study. Table 1 shows the final total magnetic moment for the different supercells.

## III. RESULTS AND DISCUSSION

### III. a. 2x2x1 supercell

The 2x2x1 supercell corresponds to the more defective material. It exhibits metallic behavior with two bands, for each polarized case, being crossed at the Fermi level. The magnetic moment arrangement is ferrimagnetic; the spin density map of Fig. 1 (a) and (b) shows the major $p_z$ character of the charge ordering. In particular the spin density is higher for atoms near the vacancy and it decays when moving away from it. Thus, the spin configuration of the system is ferrimagnetic, with higher spin-up charge in comparison to the spin-down one. As it would be expected, the $c_{eq}$ relaxed axis is longer than the experimental $c_0$ value in graphite, and it is slightly larger than the predicted one, by DFT-GGA, in pure graphite where the $c$-axis was optimized obtaining a final value of $c$=15.50 a.u.. The total magnetic moment was 0.63 $\mu_B$ for the non-relaxed cell and 2.01 $\mu_B$ for the relaxed case.

### III. b. 3x3x1 supercell

The 3x3x1 supercell presents a metallic behavior and shows magnetic response too. The energy band structure shows six up-electrons energy bands and four down-electrons energy bands that cross the Fermi level. The magnetic response could be due these localized states, what can be confirmed too by inspection of its density of states (Fig. 3 (a)). The $c_{eq}$=14.90 a.u. relaxed axis is longer than the experimental $c_0$ value in graphite, but it is slightly shorter than the predicted one by DFT-GGA in pure graphite where the *c*-axis was optimized obtaining a final value of *c*=15.50 a.u.. This trend could be in relation with some experimental findings where x-ray diffraction experiments showed that modified magnetic graphite tends to decrease the *(00l)* interlayer spacing [14,15]. Another point related with the electronic structure is the dispersion of the bands, in particular in the Γ-A direction, which is appreciable even in the case of overestimation of the *c*-axis length ($c_{eq}$= 14.90 a.u.). This band is almost filled for the spin-down electrons, showing a positive slope along Γ-A, what could be related with the tendency of contraction of the *c*-axis. This characteristic reinforces the importance of regarding this material as a real 3-D crystal. The total magnetic moment for this supercell increases from 1.76 $\mu_B$ to 2.06 $\mu_B$ when the *c*-axis is optimized. The spin arrangement is again ferrimagnetic, with a spin density which is mainly due to $p_z$ orbitals for atoms far from the vacancy (Fig. 2 (c) and (d)). An important difference with the 2x2x1 supercell case is that a strong component of $sp^2$ hybridization in the region near the vacancy is observed. These three lobules are connected symmetrically between them, with some degree of saturation. This can be deduced for the magnetic moment per vacancy which is 1.06 $\mu_B$ in the case of the $c_{eq}$, in close agreement with earlier reports [4].

### III. c. 4x4x1 supercell

The 4x4x1 supercell yields again a metallic and magnetic material. Here the system shows metallic behavior with one and two band being crossed by the Fermi level in the case of spin-up and -down electrons respectively. A difference to the previous case is that although showing some dispersion, no bands cross the Fermi level along the Γ-A path. The spin density shows similarities with the 3x3x1 case, where the ferrimagnetic order is mainly based on the $p_z$ spin density far from the point defect, and a strong localization of charge associated to $sp^2$ orbitals in vicinity of the vacancy. For this case the total magnetic moment decreases when the *c*-axis is optimized. These moments are 1.41 $\mu_B$ and 1.21 $\mu_B$ for $c_0$ and $c_{eq}$ (15.11 a.u.) respectively. This fact could be explained in terms of the stability reached in the presence of local magnetic field. In this case there are no bands crossing the Fermi level along the Γ-A path, so the stabilization energy addressed for the *c*-axis contraction (more dispersion) it is not as remarkable as it was in the case of the 3x3x1 supercell. For this reason the $c_{eq}$ is higher than in the 3x3x1 case.

### III. d. 5x5x1 supercell

The 5x5x1 supercell is different from the other ones since it is a semi-metal, with a negligible amount of states per energy at the Fermi Level. Since the system exhibit non-magnetic response the $c_{eq}$ is close to that of the pristine graphite, 15.68 a.u. and 15.50 a.u. respectively. This is another point that reinforces the structural and electronic correlations in this magnetic system.

### III. e. LDA xc-potential for the 3x3x1 supercell

The calculations performed using LDA potential for the 3x3x1 supercell, yielded a magnetic moment of 1.40 $\mu_B$ for a graphitic structure with $c_{eq}$=12.50 a.u.. This distance is in good agreement with the one obtained for pristine graphite using the LDA xc-potential $c_{eq}$=12.55 a.u.. Even when shifting away from the equilibrium value from the *c*-axis, the magnetic behavior was still present for this supercell. The magnetic moment ranges from 1.39 to 1.42 $\mu_B$ for *c*-axis values between 12.30 to 12.70 a.u., with a positive slope. Although a cell contraction is verified in comparison with the results obtained using GGA, as expected, this fact does not imply a significant variation in the magnetic behavior of the system, which is the main focus of this manuscript.

### III. f. LCNAO for the 3x3x1 supercell and single graphene sheet

In order to check our APW+lo results we performed for the 3x3x1 supercell another *ab initio* approach using the SIESTA code [16-18], which adopts a linear combination of numerical localized atomic-orbital basis sets for the description of valence electrons and norm-conserving non-local pseudopotentials for the atomic core. The pseudopotentials were constructed using the Trouiller and Martins scheme [19] which describes the interaction between the valence electrons and atomic core. The total energy was calculated within the Perdew–Burke–Ernzerhof (PBE) form of generalized gradient approximation (GGA) [7]. The real-space grid used to represent the charge density and wavefunctions was the equivalent of that obtained from plane-wave cutoff of 300 Ry, and the atomic positions were fully relaxed using a conjugate-gradient algorithm [20] until a force of 1.6E-3 Ry/a.u., and a maximum stress component of 0.05 GPa were reached. A Monkhorst Pack grid [21] with a 20x20x19 supercell, defined in terms of the actual supercell, was selected to obtain a mesh of 3200 k-points in the full Brillouin Zone. This process was repeated for different *c*-axis values without applying BSSE (basis set superposition error) correction finding the equilibrium result at $c_{eq}$= 14.65 a.u. with a total magnetic moment in cell of 2.21 $\mu_B$, which are in close agreement with the results obtained by the APW+lo method, with $c_{eq}$=14.90 a.u. and a magnetic moment of 2.06 $\mu_B$. The density of the occupied states obtained by both codes (Fig. 3(b)) show similarities in the electronic structure, in particular for those states near the Fermi level. The final atomic positions are in close agreement and no distortion for the flat sheet geometry is observed.

In order to check the reliability of this result we performed further calculations using the SIESTA code with a single graphene sheet for the 3x3, 4x4 and 5x5 supercell configurations. The distance between layers was selected as 15.00 a.u. in order to minimize the interactions among them. Identical conditions to the 3x3x1 bulk-supercell were applied for the initialization of the calculation. The only difference arises from the number of k-points selected to sample the first Brillouin zone: 3200, 1600 and 800 for 3x3, 4x4 and 5x5 respectively. In all cases, these spin polarized calculations yielded flat sheets with a net magnetic moment in agreement to what is here reported for the 3-D stacking. These results are presented in Table 2. However, differences in the symmetry of the environment surrounding the vacancies are observed. Three-fold symmetry around the vacancies was not found for single graphene sheet but observed for the 3-D stacking in the same supercell configurations. Experimental reports, using low temperature scanning tunneling microscopy, reveal the presence of a three-fold organization around point defects created on highly ordered pyrolytic graphite, HOPG [22]. We believe that this fact shows the key role of the interlayer interaction and the need to perform studies on the basis of a 3-D graphitic network instead of single graphene sheets where the effect of the stacking is missed.

A deviation from the flatness was described in previous reports [5,11], in disagreement with our results. In those articles, a pentagon that saturates two $sp^2$ orbitals was obtained when a single graphene sheet was optimized in its atomic positions, with different codes than the ones used in this manuscript. This pentagon let one of these $sp^2$ orbitals to contribute to the net magnetic moment. The structural distortion that the system undergoes puts this atom out of the plane.

## IV. CONCLUSIONS

We have investigated the magnetic order in a 3-D graphitic system where single atom vacancies in different supercells were created. We found that the magnetism arising is accompanied by a metallic behavior and that the best configuration for such occurrence corresponds to the 3x3x1 supercell, vanishing for the 5x5x1 one.

The fact that the 5x5x1 supercell does not show a net magnetic moment, while the 5x5 graphene supercell exhibit a magnetic moment of 1.72 $\mu_B$, reinforces the idea of the magnetism arising from a complex situation in which vacancies should interact with its images along the whole crystal. These results suggest that there is still work to do in experimental nanostructuration of graphite, trying to optimize the vacancies concentration and their geometrical configuration.

Additionally to these electronic properties, there exist some structural correlations that were evidenced in early reports, in particular the contraction of the *c*-axis in the case of magnetic systems. The magnetic response found by creating single carbon vacancies in a 3-D stacking using periodic supercells ranges between 0.02 and 0.14 $\mu_B$ per carbon atom, which is higher than the one found experimentally for defective bulk graphitic specimens, 1.2 x $10^{-3}$ $\mu_B$ per carbon atom [14,15]. This fact would imply that by creating this supercell configuration, this theoretical study produces defects in a more densely way than what experiments can achieve. In order to go forward towards the theoretical simulation of what could happen when defective graphite is experimentally prepared [14-15,23], further calculations using higher order vacancies have to be performed. In this work, that will be submitted soon, double or quadruple vacancies are created using different supercells. Although the calculations involved in this study are more expensive, they will provide more realistic results due to the incorporation of higher order vacancies.

Although beyond the scope of this research, a few words could be said about helium and hydrogen irradiation on HOPG samples [23]. The existence of intrinsic magnetism in a 3-D graphitic network with regularly distributed vacancies could support the idea of the role of adsorption of hydrogen in the vacancies to enhance this magnetism, as previously discussed [5]. The adsorption of hydrogen, with eventual pinning in the defects, would have the effect of stabilizing the magnetic signal of the specimen, by preventing the annihilation of the vacancies due to the recombination of Frenkel pairs, especially at room temperature. This would imply a remarkable advantage for hydrogen irradiation, in comparison to helium, that is supported by the experimental evidence, where hydrogen irradiated samples exhibit higher magnetic signal [24]. These ideas should be tested in a future work.

As a summary, an important result for the present manuscript was that flat graphene sheets, with a well defined symmetry for the atoms surrounding the vacancy, were obtained in all the 3-D stacking cases, independently on the code or the potential used.

All these properties, verified with two different *ab initio* codes, support the potential technological value of the new nano-structured modified-graphite in order to be considered in high-tech devices as a free metal material. We believe that the present report contributes to rethink the role of the 3-D stacking of graphene sheets in graphite, and to stop disregarding this effect.

# ACKNOWLEDGEMENTS

We gratefully acknowledge PEDECIBA, CSIC, PDT (project 54/38) and DICYT – Fondo Clemente Estable, Nº 10213, (Uruguayan organizations) for financial support. Ricardo Faccio would like to acknowledge PEDECIBA for a Ph.D. grant.

# REFERENCES

[1] F. Palacio and T. Makarova, eds., Carbon-based Magnetism (Elsevier, Amsterdam, 2005).

[2] D. C. Mattis, Phys. Rev. B **71**, 144424 (2005).

[3] Y. Kobayashi, K. I. Fukui, T. Enoki, and K. Kusakabe, Phys. Rev. B 73, 125415 (2006).

[4] R. Yoshikawa Oeiras, F. M. Araujo-Moreira, M. Veríssimo-Alves, R. Faccio, H. Pardo and A. W. Mombrú. *Submitted to Phys. Rev. B. (lt10700.)*

[5] P. O. Lehtinen, A. S. Foster, Yuchen Ma, A. V. Krasheninnikov and R. M. Nieminen, Phys. Rev. Lett. **93**, 187202 (2004).

[6] P. Blaha, K. Schwarz, G.K.H. Madsen, D. Kvasnicka, J. Luitz, J. (2001). WIEN2k, an Augmented Plane Wave + Local Orbitals. Program for Calculating Crystal Properties. Vienna University of Technology.

[7] (a) J. P. Perdew, K. Burke, and M. Ernzerhof, Phys. Rev. Lett. **77**, 3865 (1996); (b) J. P. Perdew, K. Burke, and M. Ernzerhof, Phys. Rev. Lett. **78**, 1396 (1997).

[8] V. N. Strocov, P. Blaha, H. I. Starnberg, M. Rohlfing, R. Claessen, J.-M. Debever, J.-M. Themlin, Phys. Rev. B **61**, 7, 4994 (2000).

[9] E. Konstantinova, S. O. Dantas, P. M. V. B. Barone, Phys. Rev. B **74**, 035417 (2006).

[10] Y. Zhang, S. Talapatra, S. Kar, R. Vajtai, S. K. Nayak and P. M. Ajayan, Phys. Rev. Lett. **99**, 107201 (2007).

[11] F. Tournus, J-C Charlier, P. Mélinon, *The Journal of Chemical Physics* 122, 094315 (2005).

[12] A. A. El-Barbary, R. H. Telling, C. P. Ewels, M. I. Heggie and P. R. Briddon, Phys. Rev. B **68**, 144107 (2003).

[13] In all the cases the supercells the *a* and *b* parameters were generated as multiple of $a_0$ and $b_0$, the in-plane geometry of the cells were retained as a hexagonal. Only the atomic positions and *c*-axis were optimized.

[14] H. Pardo, R. Faccio, F.M. Araújo-Moreira, O.F. de Lima and A.W. Mombrú. *Carbon*, **44**, 3, 565-569 (2006).

[15] A. W. Mombrú, H. Pardo, R. Faccio, O. F. de Lima, E. R. Leite, G. Zanelatto, A. J. C. Lanfredi, C. A. Cardoso, and F. M. Araujo-Moreira, *Phys. Rev.* B **71**, 100404(R) (2005).

[16] P. Ordejón, E. Artacho, J.M. Soler, *Phys. Rev.* B **53**, R10441 (1996).

[17] D. Sánchez-Portal, P. Ordejón, E. Artacho, J.M. Soler, *Int. J. Quantum Chem.* **65**, 453 (1997).

[18] J.M. Soler, E. Artacho, J.D. Gale, A. García, J. Junquera, P. Ordejón, D. Sánchez-Portal, *J. Phys.: Condens. Matter* **14** , 2745 (2002).

[19] N. Troullier, J.L. Martins, *Phys. Rev.* B **43**, 1993 (1991).

[20] W.H. Press, B.P. Flannery, S.A. Teukolsky, W.T. Vetterling, New Numerical Recipes, (Cambridge University Press, New York, 1986).

[21] H. J. Monkhorst and J. D. Pack, Phys. Rev. B **13**, 5188 (1976).

[22] J. G. Kushmerick, K. F. Kelly, H.–P. Rust, N. J. Halas, P. S. Weiss, J. Phys. Chem B, **103**, 10, 1619 (1999).

[23] P. Esquinazi, D. Spemann, R. Höhne, A. Setzer, K.H. Han, T. Butz, Phys. Rev. Lett. **91**, 227201 (2003).

# Figures Captions

**Fig. 1.** Sketch for the 2x2x1 supercell (a), 3x3x1 (b), 4x4x1 (c) and 5x5x1 (d) showing de A-B stacking of layers. (Coloured only in online version)

**Fig. 2.** (a) Spin density map for the 2x2x1 supercell in the range $\delta\rho_\sigma=0.070, -0.012\ \sigma/au^3$ (b) respective Iso-surface evaluated at $\delta\rho_\sigma=\pm 5.0E-3\ \sigma/a.u.^3$, (c) spin-density for 3x3x1 supercell in the range $\delta\rho_\sigma=0.052, -0.005\ \sigma/a.u.^3$, (d) Iso-surface for 3x3x1 supercell for $\delta\rho_\sigma=\pm 2.5E-3\ \sigma/a.u.^3$, (e) spin-density for 4x4x1 supercell in the range $\delta\rho_\sigma=0.069, -0.002\ \sigma/a.u.^3$, (f) Iso-surface for 4x4x1 supercell for $\delta\rho_\sigma=\pm 8.0E-4\ \sigma/a.u.^3$. The spin density is expressed as $\delta\rho_\sigma(\vec{r})=\rho_\uparrow(\vec{r})-\rho_\downarrow(\vec{r})$. The most positive values are represented in blue, while the most negative are in red according to each range (Coloured only in online version).

**Fig. 3.** (a) Density of states for the 2x2x1, 3x3x1, 4x4x1 and 5x5x1 relaxed supercell. DOS is expressed as states/eV/number of atoms in supercell. (b) Comparison between LCNAO (SIESTA code) and APW+lo (WIEN2k code) results for the 3x3x1 supercell.

**Fig. 4.** Spin-up (a) and -down (b) electrons energy band structures for the 3x3x1 supercell.

# Table Caption

**Table 1.** Total magnetic moments for the different supercells, relaxed (RC) and non-relaxed (NRC), indicating in each case the number of k-points in the entire first Brillouin Zone (N-kpts).
\* APW+lo results for the 3x3x1 supercell using LDA as a xc-potential
\*\* LCNAO (SIESTA code) results for the case of 3x3x1 supercell.

**Table 2.** Total magnetic moments for the different supercells indicating the number of k-points selected, in the single graphene sheet calculations.

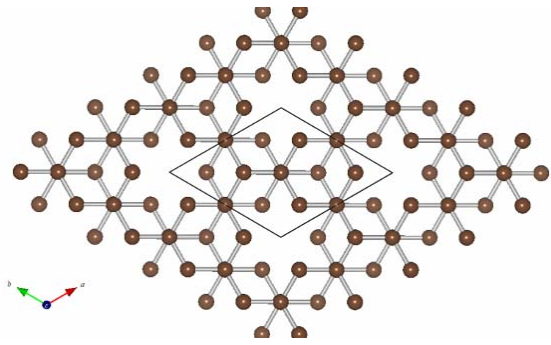

(a)

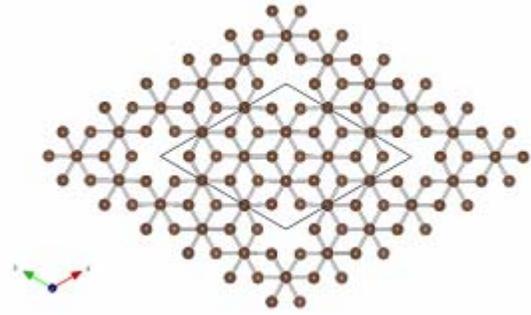

(b)

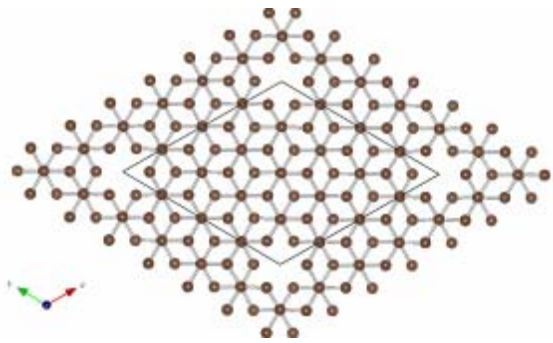

(c)

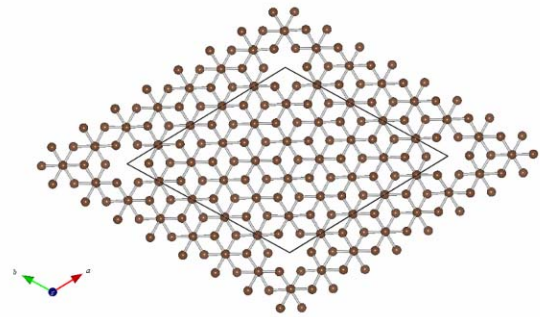

(d)

Figure 1

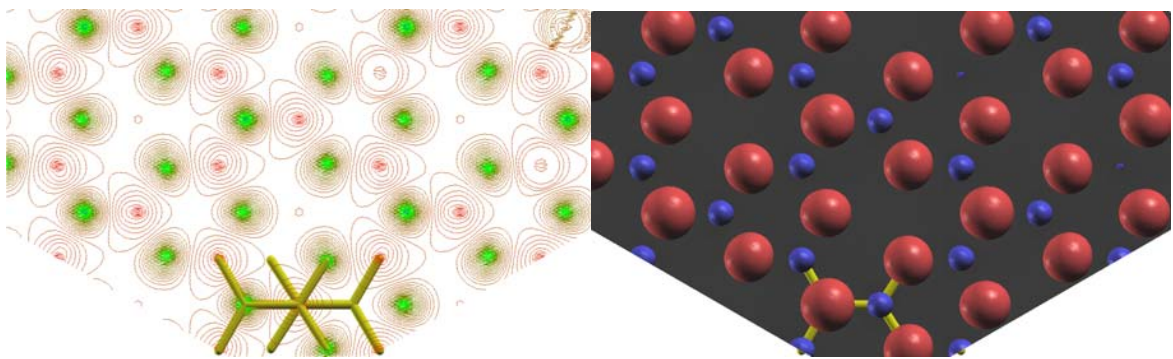

(a)  (b)

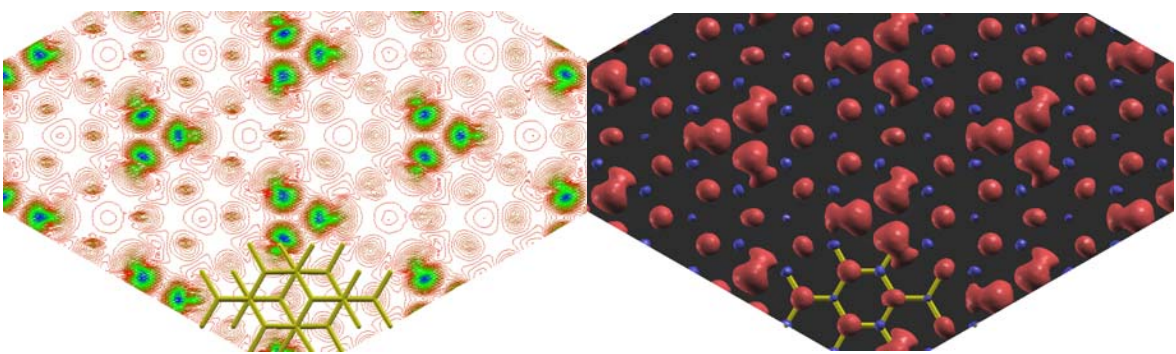

(c)  (d)

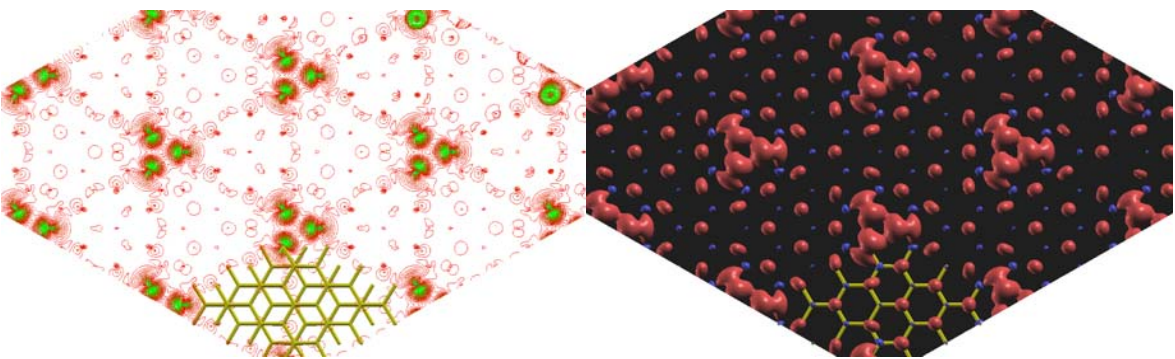

(e)  (f)

Figure 2

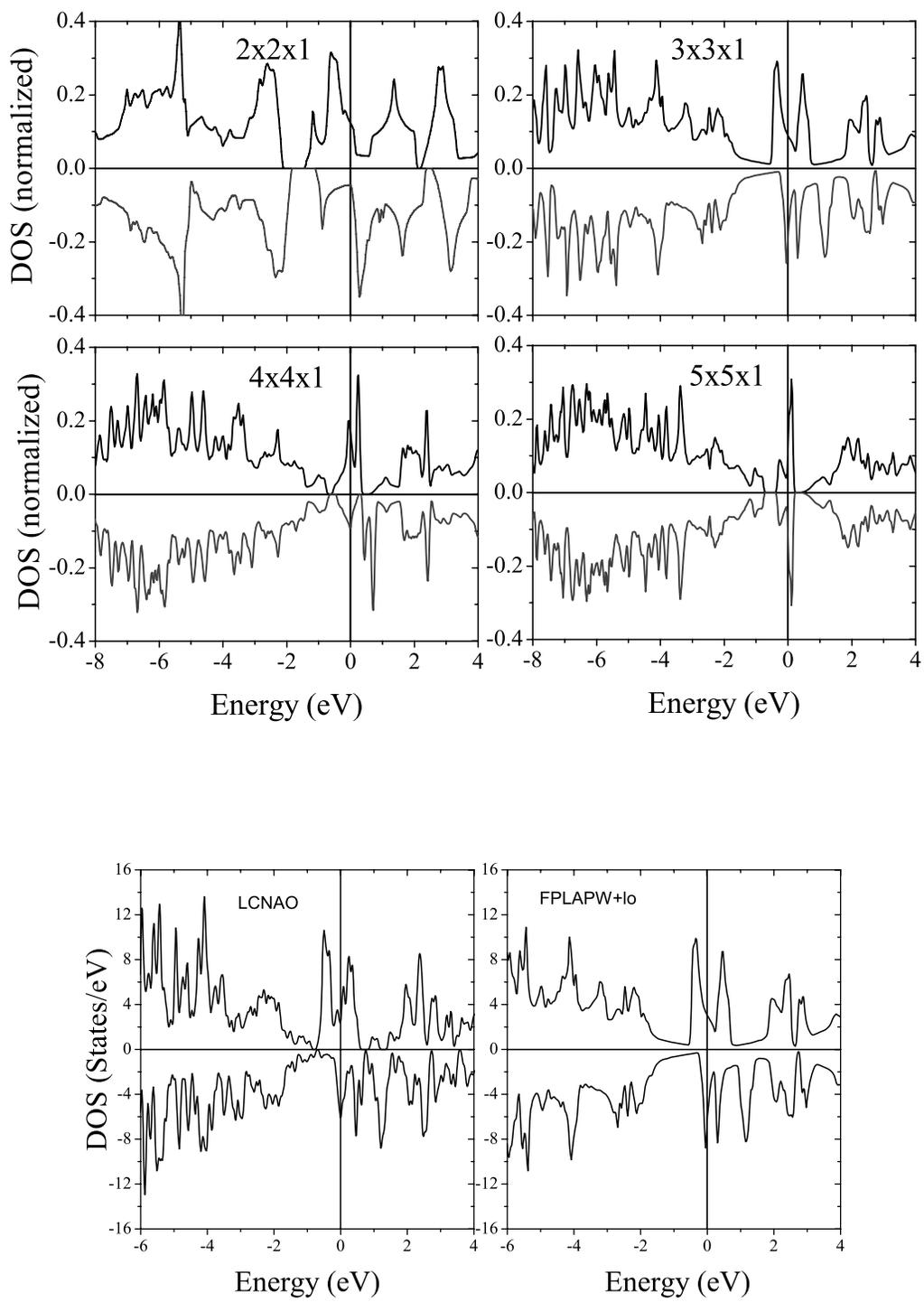

Figure 3

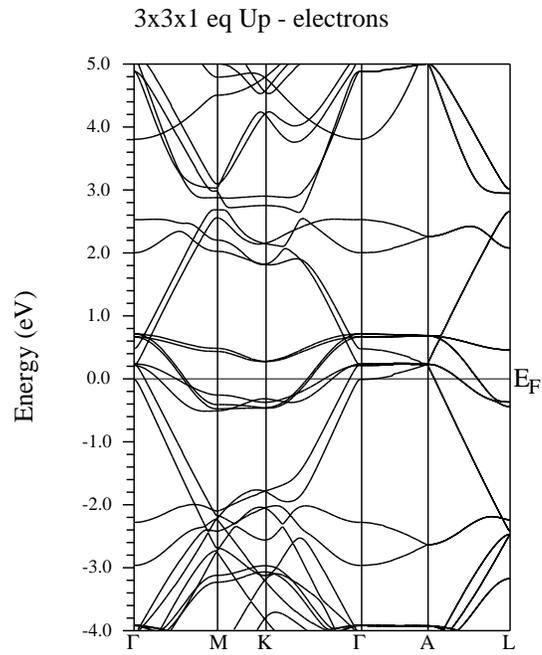

**(a)**

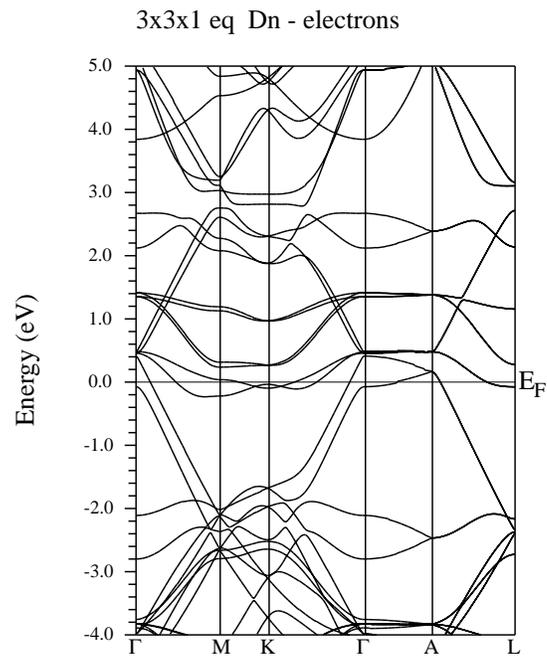

**(b)**

Figure 4

| Supercell SC | Number of atoms | Relaxed $c$ (a.u.) | N-kpts for NRC | MM for NRC ($\mu_B$) | MM for RC ($\mu_B$) | N-kpts for NRC |
|---|---|---|---|---|---|---|
| 1x1x1 | 4 | 15.50 | 16000 | 0.00 | 0.00 | 16000 |
| 2x2x1 | 14 | 16.00 | 16000 | 0.63 | 2.01 | 16000 |
| 3x3x1 | 34 | 14.90 | 3200 | 1.76 | 2.06 | 3200 |
| 3x3x1* | 34 | 12.50 | ---- | ---- | 1.40 | 3200 |
| 3x3x1** | 34 | 14.65 | ---- | ---- | 2.21 | 3200 |
| 4x4x1 | 62 | 15.11 | 1600 | 1.41 | 1.21 | 800 |
| 5x5x1 | 98 | 15.68 | 400 | 0.00 | 0.00 | 400 |

Table 1

| Supercell SC | Number of C atoms | N-kpts | MM ($\mu_B$) |
|---|---|---|---|
| 3x3 | 17 | 3200 | 2.00 |
| 4x4 | 31 | 1600 | 1.24 |
| 5x5 | 49 | 800 | 1.72 |

Table 2